\documentclass[11pt,prd,preprint,superscriptaddress,showkeys]{revtex4}
\usepackage{dcolumn}
\usepackage{bm}
\input epsf

\begin{document}


\title{{\bf { Presence of anisotropic  pressures in Lema\^{i}tre-Tolman-Bondi cosmological models}}}

\author{Sergio del Campo}

\affiliation{Instituto de F\'isica, Pontificia Universidad
Cat\'olica de Valpara\'iso, Av. Brasil 2950, Valpara\'iso, Chile.}

\author{V\'ictor H. C\'ardenas}

\affiliation{Departamento de F\'{\i}sica y Astronom\'ia, Universidad
de Valpara\'iso, A. Gran Breta\~na 1111, Valpara\'iso, Chile}
\author{Ram\'{o}n Herrera}
\affiliation{Instituto de F\'isica, Pontificia Universidad
Cat\'olica de Valpara\'iso, Av. Brasil 2950, Valpara\'iso, Chile.}

\date{\today}

\begin{abstract}
We describe Lema\^{i}tre-Tolman-Bondi cosmological models where an anisotropic  pressures is considered.
By  using recent astronomical observations coming from supernova of Ia types
we constraint the values of the parameters that characterize our models.
\end{abstract}


\maketitle

\section{Introduction}
About a decade ago, current local measurements of redshift and
luminosity-distance relations of Type Ia Supernovae
(SNIa)\cite{S10} indicate that the expansion of the universe
presents an accelerated phase \cite{R98,P99}. In fact, the
astronomical measurements showed that the SNe at a redshift of $z
\sim 0.5$ were systematically fainted, which could be attributed
to an acceleration of the universe. These local observations can
then be extrapolated to the universe at large, and the resulting
Friedmann-Robertson-Walker (FRW) model can be used to interpret
all cosmological measurements, which strongly support a dominant
dark energy component, responsible for this cosmological
acceleration.

It is usual to say that this accelerated expansion is due to a
cosmological constant, the so called concordance $\Lambda$CDM,
which fits the observations quite well, but there is no
theoretical understanding of the origin of this cosmological
constant or its magnitude (the well known cosmological constant
problem). Also, this model presents a coincidence problem of late
cosmic acceleration, which establishes the fact that the density
values of dark matter and dark energy are of the same order
precisely today\cite{dCHP2008}.

In order to get rid of the above problems there has been put
forward other kind of different models, where some exotic, unknown
and uncluttered matter component, dubbed dark energy\cite{HT99}
(see also Refs. \cite{T10,C09} for recent reviews) is considered.
Among these possibilities are quintessence \cite{C98,Z98},
k-essence \cite{Ch00,AP00,AP01}, phantom field \cite{C02,C03,H04},
holographic dark energy \cite{Z07,W07,julio}, etc. (see
Ref.~\cite{D09} for model-independent description of the
properties of the dark energy and Ref.~\cite{S09} for possible
alternatives). However, although fundamental for our understanding
of the evolution of the universe, the nature of this quintessence
remains a completely open question nowadays, moreover, very often
they require fine-tuning\cite{CST06,S06}.

All these models work under the assumption of the Cosmological
Principle, which imposes homogeneity and isotropy. But, the question
is, are these symmetries consistent with observations? We know that
inhomogeneities are abundant in the universe: there are not only
clusters of galaxies but also large voids. It has usually been
argued that these symmetries should only be valid on very large
scales.

It is true that this recognition is a result of the strict use of
a FRW homogeneous and isotropic model. However, inhomogeneous
models have been put forward. For instance, if phantom energy is
considered, then it is possible to sustain traversable wormholes
in which the phantom matter is an inhomogeneous and anisotropic
fluid\cite{CLdCCS08,CdCMS09,CdC12}. Also, the authors of Ref.
\cite{GGMP} considered inhomogeneous metrics, but with an equation
of state given by $P=\omega \rho$, with $\omega$ the equation of
state parameter being constant. Here, they study the gravitational
collapse of spherical symmetric physical systems. On the same line
of reasoning, the author of Ref. \cite{LL06} studied a
Lem\^{a}tre-Tolman-Bondi (LTB) model, taking into account a
perfect fluid, in which they describe either a stellar or galactic
system, where the collapse of these systems might yield to a
possible formation of a black hole.  Also, the effect of pressure
gradients on cosmological observations by deriving the luminosity
distance – redshift relations in spherically symmetric,
inhomogeneous space-times endowed with a perfect fluid was
considered in Ref.\cite{Lasky:2010vn}.

If inhomogeneities are properly considered it might be possible to
explain the observations without introducing the dark energy
component. Recently, this possibility has renewed interest in
inhomogeneous cosmological models, especially in the
 LTB solution\cite{L33,T34,B47}, which
represents a spherically symmetric dust-filled universe. Because
of its simplicity this solution has been considered to be most
useful to evaluate the effect of inhomogeneities in the observable
universe, like the luminosity distance-redshift
relation\cite{C00,MHE97,INN02}. In most of these kind of models
the matter component has been considered to be a perfect fluid
with vanishing equation of state parameter, i.e. $\omega=0$,
corresponding to a dust fluid.

 Recently, numerical simulations of large scale structure
evolution in an inhomogeneous LTB model of the Universe was
studied in Ref.\cite{P1} and the LTB model whose distance-redshift
relation agrees with that of the concordance $\Lambda$CDM model in
the whole redshift domain and which is well approximated by the
Einstein-de Sitter universe at and before decoupling and the
matching peak positions in the CMB spectrum was considered in
Ref.\cite{P2}. Also, the kinematic Sunyaev-Zel'dovich (kSZ) effect
and  the CMB anisotropy observed in the rest frame of clusters of
galaxies was developed in Ref.\cite{P3}.


In this work we want to address the study of an inhomogeneous LTB
model, in which the matter component will be an inhomogeneous and
anisotropic fluid \cite{CLdCCS08,CdCMS09}, where the radial and
tangential pressures are given by $p_r=\omega \rho$ and
$p_\bot=\omega_\bot \rho$, respectively. As far as we know no one
has studied this problem previously.


The outline of the paper is as follows. The next section presents a
short review of  the Lema\^{i}tre-Tolman-Bondi model.  In the
Sections III and IV we discuss the analytical solutions for our
model, respectively. Here, we give explicit expressions for the
functions of the radial coordinate and the luminosity distance.
Finally, our conclusions are presented in Section V.  We chose units
so that $c=\hbar=1$.



\section{The Lema\^{i}tre-Tolman-Bondi Models}
Lema\^{i}tre-Tolman-Bondi metric can be written as
Refs.\cite{L33,T34,B47}
\begin{equation}\label{metric}
ds^2=dt^2-\frac{A'(r,t)^2}{1-k(r)}dr^2-A(r,t)^2d\Omega^2,
\end{equation}
where $A$ is a function of the radial coordinate $r$ and the time
coordinate $t$, i.e., $A(r,t)$,  $k(r)$ is an arbitrary function
of the coordinate $r$, $A'(r,t)=\partial A/\partial r$ and
$d\Omega^2=d\theta^2+\sin^2\theta d \phi^2$.

Let us suppose that the universe is filled with  fluid with a
stress energy tensor as $T_{\mu \nu}=diag(\rho,
p_{\perp},p_{\parallel},p_{\parallel})$, where $\rho(r,t)$,
$p_{\perp}$ and $p_{\parallel}$ are respectively the energy
density, the radial pressure $p_{\perp}(r,t)=p_r(r,t)$  and
lateral pressure $p_{\parallel}(r,t)=p_\phi(r,t)=p_\theta(r,t)$ as
measured by observers who always remain at rest at constant $r$,
$\phi$ and $\theta$.

From the Einstein equations we obtain for the $00$-component, we
get
\begin{equation}\label{00}
2\frac{\dot{A}}{A}\frac{\dot{A'}}{A'}+\frac{k(r)}{A^2}
+\frac{\dot{A}^2}{A^2}+\frac{k'(r)}{AA'}=8\pi G \rho.
\end{equation}
The other components of the Einstein equations; the $rr$ equation
gives
\begin{equation}\label{rr}
2\frac{\ddot{A}}{A}+\frac{k(r)}{A^2}+ \frac{\dot{A}^2}{A^2}=-8\pi G
p_{\perp},
\end{equation}
and from the $\theta \theta$ and $\phi \phi$ equations are
\begin{equation}\label{2teta}
\frac{\partial}{\partial r}\left( 2\ddot{A} A + k(r) + \dot{A}^2
\right)=-8\pi G p_{\parallel} \frac{\partial}{\partial r}(A^2),
\end{equation}
where the dot denoting the partial derivative with respect to $t$.

Combining both equations (\ref{rr}) and (\ref{2teta}) we obtain
\begin{equation}\label{cons}
\frac{\partial p_{\perp}}{\partial r}A^2 +
(p_{\perp}-p_{\parallel})\frac{\partial}{\partial r}A^2=0.
\end{equation}

From  the conservation equation $T^\mu_{\;\;\;\nu\,;\mu}=0$, we
have that
\begin{equation}\label{ropto}
\frac{\partial \rho}{\partial t} +
\frac{\dot{A'}}{A'}(p_{\perp}+\rho)+2H(\rho+p_{\parallel})=0,
\end{equation}
where $H \equiv \dot{A}/A$. Note that if we restrict  ourself to
the case $A(r,t)=a(t)r$, these equations are exactly those
obtained in Ref.\cite{CdCMS09}, in which $ \frac{\partial
\rho}{\partial t} +  H(3\rho+p_{\perp} +2p_{\parallel})=0, $ with
the identifications $p_r=p_{\perp}$ and $p_l=p_{\parallel}$. Note
also that when $p_{\perp}=p_{\parallel}$, Eq.(\ref{cons}) result
in that  $p_{\perp}$ is only function of time, and leads to the
usual conservation equation.

Now we shall require that the radial  and the lateral pressures
have barotropic equations of state. Thus, we can write
$p_{\perp}=\omega_{\perp}\rho$  and
$p_{\parallel}=\omega_{\parallel}\rho$,  where $\omega_{\perp}$
and $\omega_{\parallel}$ are constant eq. of state parameters.
Note that in the  case when $\omega_{\perp}=\omega_{\parallel}=0$,
we obtained the standard LTB model where the matter source have a
negligible pressure, representing dust.


Introducing the function
\begin{equation}\label{fdert}
F(r,t) \equiv A (\dot{A}^2 + k(r)),
\end{equation}
equations (\ref{00}) and (\ref{rr}) can be written in a compact form
\begin{eqnarray}
  F' &=& 8\pi G \rho A' A^2, \\
  \dot{F} &=& -8\pi G \omega_{\perp} \rho \dot{A} A^2.
\end{eqnarray}
The function expressed by Eq.(\ref{fdert}) is the generalization
of the one used in \cite{Enqvist:2006cg}, \cite{Enqvist:2007vb}
and \cite{GBH}, where the dust version were studied. We can
recover this case by explicitly taking $\omega_{\perp}=0$ in the
second equation, which implies that $F$ is only a function of $r$.

In general, equation (\ref{fdert}) could  be considered as a
Friedmann Equation for a non-zero pressure case. Thus, we rewrite
\begin{equation}\label{friedman}
H^2(r,t)=H_0^2(r) \left[ \Omega_M(r) \left(
\frac{A_0}{A}\right)^{3(1+\gamma)}+(1-\Omega_M(r))\left(
\frac{A_0}{A}\right)^{2} \right],
\end{equation}
where we have defined as before $H \equiv \dot{A}/A$ and $\gamma$ is
a parameter that is zero for the dust case. The matter density and
spatial curvature are defined by
\begin{eqnarray}
  \Omega_M(r) &=& \frac{F_0(r)}{A_0^3 H_0^2}, \label{omr} \\
  \Omega_M(r)-1 &=& \frac{k(r)}{A_0^2 H_0^2}, \label{kder}
\end{eqnarray}
where the subscripts $0$ correspond to present values,
$F_0=F(r,t_0)$, $H_0=H(r,t_0)$ and $A_0=A(r,t_0)$. These relations
are very important because, once we perform the analysis of our
models with the observational data, we can actually \textit{find}
the profile $\Omega_M(r)$ that best fist the data. This is done
explicitly  in the next section.


For an observer located at $r=0$, incoming light travels along
radial null geodesics, i.e.  $ds^2=d\Omega^2=0$, so time decreases
with, $dt/dr<0$, and thus we have
\begin{equation} \label{dtdr}
\frac{dt}{dr}=-\frac{A'(r,t)}{\sqrt{1-k(r)}},
\end{equation}
which together with the redshift equation (see Ref.\cite{GRC})
\begin{equation} \label{drdz}
\frac{dr}{dz}= \frac{\sqrt{1-k(r)}}{(1+z)\dot{A'}(r,t)},
\end{equation}
enable us to write down the relation between redshift $z$ and time
\begin{equation} \label{dtdz}
\frac{dt}{dz}=- \frac{A'(r,t)}{(1+z)\dot{A'}(r,t)}.
\end{equation}
Equations (\ref{drdz}) and (\ref{dtdz}) are used to find the
functions $t(z)$ and $r(z)$. From these expressions, one can
immediately write down  the luminosity distance, the co-moving
distance and the angular diameter distance as a function of
redshift \cite{GBH}
\begin{eqnarray}
d_L(z)& = & (1+z)^2A(r(z),t(z)), \label{dlz} \\
d_C(z)& = & (1+z)A(r(z),t(z)), \label{dcz} \\
d_A(z)& = & A(r(z),t(z)), \label{daz}
\end{eqnarray}
where $z$ and $A$ are evaluated along the radially-inward moving
light ray.

\section{Separation of variable}
In the following we will derive analytic solutions for our LTB
model, from separation of variable of the $\rho$ and $A$,
respectively. From Eq.(\ref{cons}) we find that
\begin{equation}
\rho(r,t)=
f(t)A^{-2(1-\frac{\omega_{\parallel}}{\omega_{\perp}})},
\end{equation}
where $f(t)$ is an arbitrary function of $t$. This suggest us that
we can look for solutions under the following assumption:
$\rho=\rho_1(t)\rho_2(r)$, and also $A(r,t)=A_1(t)A_2(r)$.

Inserting these identities in to Eqs.(\ref{cons}) and
(\ref{ropto}) we find that
\begin{eqnarray}
\rho_1(t)& = & C_1 A_1^{-\alpha}, \hspace{1cm} \alpha=3+2\omega_{\parallel}+\omega_{\perp}, \label{ro1} \\
\rho_2(r)& = & C_2 A_2^{-\beta},
\hspace{1cm}\beta=2\left(1-\frac{\omega_{\parallel}}{\omega_{\perp}}\right).
\label{ro2}
\end{eqnarray}

Now, from Eqs.(\ref{00}) and (\ref{rr}) we find  the following
differential equation for the arbitrary function $k(r)$ as a
function of $A_2(r)$
\begin{equation} \label{master}
2\omega_{\perp} \frac{dk}{dA_2^2}+ \frac{k}{A_2^2}(1+
\omega_{\perp})=-\dot{A_1}^2(1+3 \omega_{\perp}) - 2
\ddot{A_1}A_1=C.
\end{equation}
As we can see, the left hand side depends exclusively of functions
depending on the variable $r$ and the right hand side everything
depends on $t$, so both sides are equal to a constant $C$. We will
have different family of solutions depending on the value that the
constant $C$ could take. Notice also that $\omega_{\parallel}$ is
not relevant parameter in obtaining expressions for $A_1$ and
$A_2$.


Using this separation of variables, the metric (\ref{metric}) takes
the form
\begin{equation}\label{metricmod}
ds^2 = dt^2 - A_1^2(t) \left[ \frac{dA_2^2(r)}{1-k(r)} - A_2^2(r)
d\Omega^2 \right],
\end{equation}
where is clear that $A_2(r)$ can be considered as a new radial
coordinate. In this case, instead of trying to get the explicit
functional form $r(z)$, we just need to obtain $A_2(z)$. Notice
that although Eq.(\ref{metricmod}) looks very similar to the FRW
one, the freedom in choosing $k(r)$ keep the model inhomogeneous.
However, if we assume in Eq.(\ref{master}) that
$\omega_{\perp}=0$, then we get that $k=C A_2^2$, and thus  it
reduces Eq.(\ref{metricmod}) to the FRW homogeneous metric
solution. It means that this separation of variables,  reduces
directly to the FRW model in the case of the universe it filled by
dust.


At this point, we can immediately write down the explicit expression
for $A_1$, in terms of the redshift $z$. From Eq.(\ref{dtdz}), we
get
\begin{equation} \label{a1dz}
\frac{d
t}{dz}=-\frac{A_1}{(1+z)\,\dot{A}_1}\;\;\Rightarrow\;\;\;A_1(z) =
\frac{A_1(0)}{1+z},
\end{equation}
with $A_1(0)=A_1(z=0)$. Actually, this result is independent of
any time dependance that we can anticipate for $A_1(t)$. This is
somehow the analog to the FRW relationship between the scale
factor $a(t)$ and the redshift $z$.

On the other hand, in general terms it is not possible to write
down an explicit expression for $A_2(z)$. However, we could write
down Eq.(\ref{drdz}) under the assumptions set of  the previous
section, leading to the equation
\begin{equation} \label{a2dr}
\frac{A'_2 dr}{\sqrt{1-k(r)}}=\frac{d \log(1+z)}{\dot{A_1}}.
\end{equation}
In order to solve this equation, we need the explicit expressions
for $A_1(t)$, and a relationship between $k(r)$ and $A_2(r)$. Both
relations follows from Eq.(\ref{master}). In the next section, we
consider several cases where this procedure enable us to write
down an explicit expression for $A_2(z)$, which together with
Eq.(\ref{a1dz}) enable us to compute the luminosity distance given
by Eq.(\ref{dlz}), and thus, we could test each model by using
supernova data.


In the following we will study different families of solutions,
which are characterized by the choice of  the constant $C$ (see
Eq.(\ref{master})). In each case, we perform a Bayesian analysis
using supernova data to constraint the parameters appearing  in
the model. If the corresponding scenario is favored by the
observations, then we can use the results given by Eqs.(\ref{omr})
and (\ref{kder}), and plot the profile of the function
$\Omega_M(r)$.

By using the separation of variables described at the beginning of
this section,  we found that $H(r,t)=\dot{A_1}/A_1$ is  a function
of time only, so that $H_0$ is a constant independent of the $r$
coordinate. Also, if we take into account Eq.(\ref{a1dz}), we find
that $A_0(r)=A_1(z=0)A_2(r)$, where again it is clear that $A_2$
can be considered as a new radial coordinate. In the following, we
proceed  in order to find some specific solutions to our model.

\section{ Some specific solutions}\label{sol}

The simplest case to solve Eq.(\ref{master}) is when  the constant
$C$ takes the value zero. Here, we obtain that
\begin{eqnarray}
A_1(t)& = & c_1 [3(1+\omega_{\perp})t ]^{\frac{2}{3(1+ \omega_{\perp})}} , \label{a1} \\
A_2(r)& = & c_2 k(r)^{-\frac{\omega_{\perp}}{1+ \omega_{\perp}}},
\label{a2}
\end{eqnarray}
where $c_1$ and $c_2$ are two arbitrary constants. The solution
 expressed by Eq.(\ref{a1}) is obtained imposing the requirement that $A_1(t=0)=0$.
Combining Eqs.(\ref{a1dz}) and (\ref{a1}) we can find a relation
between the time $t$ and the redshift $z$ given by
\begin{equation} \label{tdz}
\left(\frac{t_0}{t}\right)^{\frac{2}{3(1+\omega_{\perp})}}=1+z,
\end{equation}
where $t_0$ expressed the present times.


On the other hand, from Eq.(\ref{a2}) we get that $k(A_2)=c
A_2^{-(1+ \omega_{\perp})/\omega_{\perp}}$, and from
Eq.(\ref{kder}),  we obtain the following expression for the
density parameter
\begin{equation}\label{omder2}
\Omega_M(A_2)=1+\frac{c}{A^2_1(0)H_0^2} A_2^{-(1+
3\omega_{\perp})/\omega_{\perp}}.
\end{equation}
Note that $A_2$ could be considered as a new radial coordinate.
This latter expression, allows us to plot the radial profile of
the inhomogeneous density matter. Notice also that a ``void''
solution has to satisfy that as $A_2\rightarrow \infty$,
$\Omega_M$ has to \textit{increase} towards an asymptotic constant
value. This means that the second term in Eq.(\ref{omder2}) have
to decrease as $A_2\rightarrow \infty$. This is possible only if
$\omega_{\perp}<-1/3$ or $\omega_{\perp}>0$.


In order, to obtain $A_2(z)$ we need to integrate Eq.(\ref{a2dr}).
In this way, the integral in the left hand side can be written as
\begin{equation}\label{inta}
\int \frac{A'_2 dr}{\sqrt{1-k(r)}}=\int \frac{dA_2}{\sqrt{1-cA_2^{\alpha}}},
\end{equation}
where $\alpha=-(1+\omega_{\perp})/\omega_{\perp}$. This latter
integral  can be done analytically only in some cases where the
parameter $\omega_{\perp}$, takes some specific values.

\subsection{Case $\omega_{\perp}=-\frac{1}{2}$.}\label{subsA}

Here, we have $\alpha=1$ and the integral of the Eq.(\ref{inta})
can be done directly. Inserting this expression in to
Eq.(\ref{a2dr}), and having in mind the Eqs.(\ref{a1dz}) and
(\ref{a1}), we get
\begin{equation}\label{a2dz}
A_2(z)=\frac{1}{c}\left[ 1-\left( \sqrt{1-c A_2(0)}-
\frac{3t_0c}{2A_1(0)}\left\lbrace (1+z)^{1 \over 4}-1
\right\rbrace \right)^2\right].
\end{equation}
The luminosity distance given by Eq.(\ref{dlz}) is then computed
using the Eqs.(\ref{a1dz}) and (\ref{a2dz}). This expression has
three free parameters; $e_1=A_1(0)/c$, $e_2=cA_2(0)$ and
$e_3=3t_0/4$. To test our model with observational data we use the
Supernova Cosmology Constitution sample \cite{constitution},
consisting of 397 SNIa expanded in a redshift range $0.015<z<1.5$
and the more recent Union2 sample \cite{Union2}, consisting of 557
SNIa.

\begin{figure}[h]
\centering \leavevmode\epsfysize=5cm \epsfbox{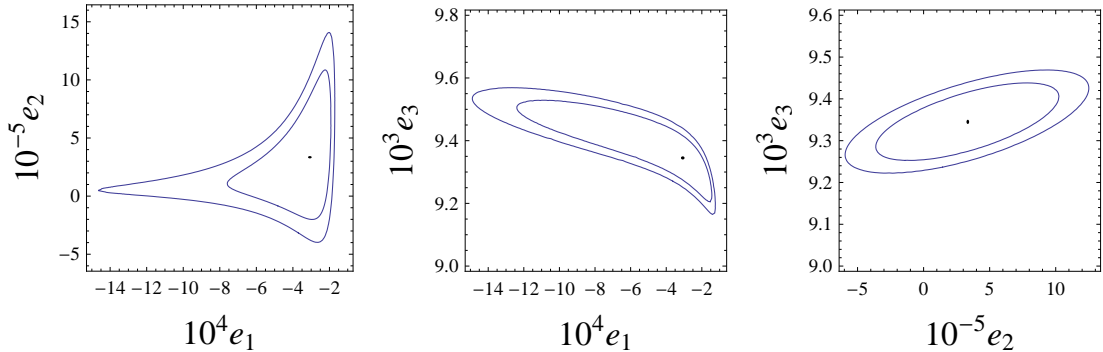}\\
\caption{\label{fit1} Here, it is shown the result of the bayesian
analysis of the model. We use the Constitution
(\cite{constitution}) data set consisting in 397 SNIa. The best
fit value for our parameters $e_1=A_1(0)/c$, $e_2=cA_2(0)$ and
$e_3=t_0$ is shown with confidence contours showing the $68.97\%$
and $90\%$ regions. }
\end{figure}

In Fig. \ref{fit1} we see the confidence contours for the three
parameters under consideration using the Constitution set. The best
fit value  for our parameter $A_1(0)/c$ is shown with two horizontal
lines indicating the range for $68.97\%$ (continuous line) and
$90\%$ (dashed line) confidence region.

\begin{figure}[h]
\centering \leavevmode\epsfysize=5cm \epsfbox{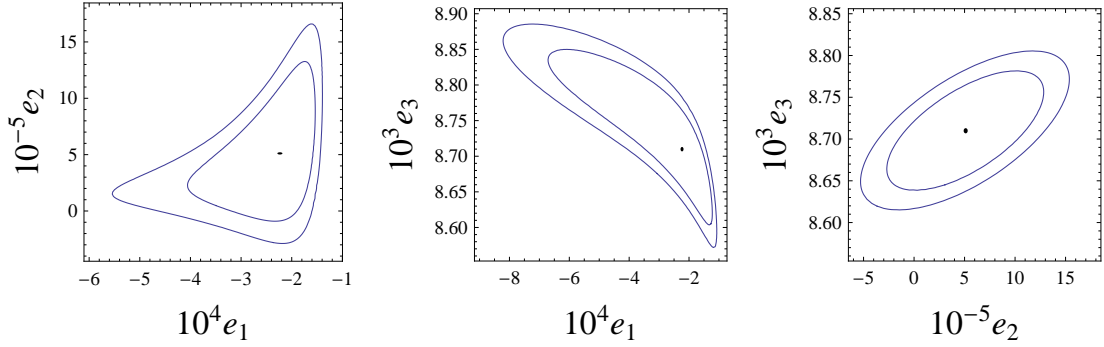}\\
\caption{\label{fit2} The figure  shows the result of the bayesian
analysis of the model. We use the Union 2 (\cite{Union2}) data
set. The best fit value for our parameters $e_1=A_1(0)/c$,
$e_2=cA_2(0)$ and $e_3=t_0$ is shown with confidence contours
showing the $68.97\%$ and $90\%$ regions. }
\end{figure}

Using the numerical value for the parameters and the relation
(\ref{kder}) we can find an explicit expression for the density
parameter as a function of $A_2$. Clearly the solution with
$\omega_{\perp}=-1/2$ satisfy the requirement for a void solution,
as we mention in the last paragraph. The results of the numerical
analysis gives a very small value for $A_2(0)$, which does not
permit us to estimate the value for $\Omega_M(z=0)$. The same
happens using the Union 2 data set \cite{Union2}. The results of
the fitting process is shown in Fig.(\ref{fit2}). In the first
case $\chi^2_{red}=1.18$ and in the second one
$\chi^2_{red}=0.98$, showing that this is a quite good model.

\subsection{Case $\omega_{\perp}=-\frac{1}{3}$.}

Here $\alpha=2$ and from Eq.(\ref{inta}), we get
\begin{equation}
A_2(z)=\frac{1}{\sqrt{c}}\,
\sin\left[\frac{\sqrt{c}}{2\,c_1}\,\log(1+z)\right]=
\frac{1}{\sqrt{c}}\,\sin\left[\frac{\sqrt{c}\,t_0}{A_1(0)}\,\log(1+z)\right].
\end{equation}
Here, the solutions will be a $\sin$ or $\sinh$function, depending
on the sign of the $c$ constant.  Although it is possible to write
down in explicit form the luminosity distance, and then test it
under a bayesian analysis, we get that $k(A_2) \propto A_2^2$, and
thus  leading to a FRW metric form and no inhomogeneity is
obtained, as can be seeing from Eq.(\ref{omder2}). Thus, in this
case we recover the homogeneous FRW model.

\subsection{Case $\omega_{\perp}=-\frac{2}{3}$.}

Another analytical solution can be obtained by using
$\omega_{\perp}=-\frac{2}{3}$. In this case $\alpha=1/2$ and the
relation $k(A_2) \propto \sqrt{A_2}$ enable us to have a
inhomogeneous universe. This also satisfies the criteria of a
``void'' solutions, because $\omega_{\perp}<-1/3$. However, the
bayesian analysis indicate that this model is not appropriated for
describing the observations of the SNIa.


Let us consider now the  case where the equations expressed by
(\ref{master}) are equals to a constant $C$ different from zero.
The general solution for $A_2(r)$ it is found to be
\begin{equation}\label{kda2}
k(A_2)=cA_2^{-\frac{1+\omega_{\perp}}{\omega_{\perp}}}+
\frac{C\,A_2^2}{1+3\omega_{\perp}},
\end{equation}
where the first term corresponds to the homogeneous solution of
Eq.(\ref{master}), i.e., $C=0$ (see Eq.(\ref{a2})). By taking into
account Eqs.(\ref{kda2}) and (\ref{kder}) we notice that this
class of solution are not permitted, due to the presence  of the
second term proportional to $A_2^2$ in Eq.(\ref{kda2}). We see
that, as $A_2 \rightarrow \infty$ the matter density parameter
$\Omega_M$ goes to $\pm \infty$, depending on the global sign of
this term (which depends on $C$ and $\omega_{\perp}$). This
behavior makes the model unacceptable.

\section{Conclusions \label{conclu}}

In this paper we have studied a certain class of inhomogeneous LTB
universes, in which the matter component is an inhomogeneous and
anisotropic fluid \cite{CLdCCS08,CdCMS09}, where the radial and
tangential pressures are given by $p_r=\omega \rho$ and
$p_\bot=\omega_\bot \rho$, respectively. We have found analytical
solutions to the system in order to compute explicitly the
luminosity distance that was tested with the SNIa observations. We
use both the Constitution \cite{constitution} and Union 2
\cite{Union2} data sets. Our equations enable us to find directly,
without any arbitrary ansatz, the profile of the density matter
$\Omega_M$ as a function of the radial coordinate. In this way,
the fitting process with observations can fix the value of the
density matter, both at $z=0$ and at $z \rightarrow \infty$. We
have found that the only model favored by the observations is of
the ``void'' type solution, where $\Omega_M$ is a function that
increase with the increment of the radial coordinate, leading to a
constant asymptotic value. Although we can not determine the
current value for the matter density, because of the very low
value for $A_2(0)$ in the fitting process (see comments at the end
of subsection \ref{sol}), we think that the method could be of
interest in the case of nearly FRW inhomogeneous models.





A necessary next step in testing these kind of models is to
consider other observational probes. At present there is no
agreement about how to proceed in the use of CMB and Large scale
structure as additional probes \cite{26,35} even in the simplest
LTB dust model. Although progress has been made \cite{cmb_dp}, we
need to understand better the evolution of density perturbations
in LTB spacetimes. Because at present we do not have a clear
framework to address these problems, we have to look for
additional observational probes. For example, in a recent work
\cite{dePutter:2012zx} the authors used galaxy ages \cite{GA}. The
results still favored the $\lambda$CDM model against the simplest
dust LTB solution, in agreement with other probes.


Note also that we have not addressed the stability issue of our
models. But, since it becomes similar to a FRW-type of model (see
expression (\ref{metricmod})) we expect that it model studied here
is effectively stable. The demonstration could be carried in a
similar way than that followed in the FRW case. However, here the
situation become more complicate due to its homogeneity. We expect
to address these points in the coming future.

\section*{Acknowledgements}

This work was funded by Comision Nacional de Ciencias y
Tecnolog\'{\i}a through FONDECYT Grants 1110230 (SdC, VHC and RH)
and 1090613 (RH and SdC), and by DI-PUCV Grant 123710 (SdC) and
123703 (RH). Also, VHC thanks to DIUV project No. 13/2009.


\end{document}